# A FLEXIBLE HIGH-BANDWIDTH LOW-LATENCY MULTI-PORT MEMORY CONTROLLER

**Xuan-Thuan NGUYEN[1], Duc-Hung LE[2, *], Trong-Tu BUI[2], Huu-Thuan HUYNH[2], Cong-Kha PHAM[1]**

[1]*The University of Electro-Communications, 1-5-1 Chofugaoka, Chofu, 182-8585 Tokyo, Japan*

[2]*University of Science, Vietnam National University – Ho Chi Minh City, 227 Nguyen Van Cu, District 5, Ho Chi Minh City, Viet Nam*

[*]Email: *ldhung@hcmus.edu.vn*

**Abstract.** Multi-port memory controllers (MPMCs) have become increasingly important in many modern applications due to the tremendous growth in bandwidth requirement. Many approaches so far have focused on improving either the memory access latency or the bandwidth utilization for specific applications. Moreover, the application systems are likely to require certain adjustments to connect with an MPMC, since the MPMC interface is limited to a single-clock and single-data-width domain. In this paper, we propose efficient techniques to improve the flexibility, latency, and bandwidth of an MPMC. Firstly, MPMC interfaces employ a pair of dual-clock dual-port FIFOs at each port, so any multi-clock multi-data-width application system can connect to an MPMC without requiring extra resources. Secondly, memory access latency is significantly reduced because parallel FIFOs temporarily keep the data transfer between the application system and memory. Lastly, a proposed arbitration scheme, namely window-based first-come-first-serve, considerably enhances the bandwidth utilization. Depending on the applications, MPMC can be properly configured by updating several internal configuration registers. The experimental results in an Altera Cyclone V FPGA prove that MPMC is fully operational at 150 MHz and supports up to 32 concurrent connections at various clocks and data widths. More significantly, achieved bandwidth utilization is approximately 93.2 % of the theoretical bandwidth, and the access latency is minimized as compared to previous designs.

## 1. INTRODUCTION

The rapid development of silicon technology in the last decade has allowed FPGAs to perform computing-intensive applications on account of a vast amount of integrated lookup tables, dedicated registers, embedded digital signal processing, and memory blocks. This was exemplified by an FPGA design that calculated the 2K×2K two-dimensional Discrete Fourier Transform in just under 26.2 ms [1]. However, system performance is more or less negatively affected while accessing external memory without efficient controller usage. Taking the example above, processing time of 26.2 ms could only be achieved if the efficiency of the memory controller is higher than 80 %. As a result, efficient memory controllers have become increasingly attractive to researchers.

To date, some simulation approaches to high-performance controllers have been proposed. E. Ipek *et al*. [2] introduced a reinforcement-learning-based controller that optimized the scheduling policy on the fly by observing the current and previous system states, thus improving the bandwidth utilization by 22 % compared to the original controllers. A prefetch-aware controller from C. J. Lee *et al*. [3] minimized the number of redundant prefetches so as to reduce the extra bandwidth consumption by 10.7 % and 9.4 % on four and eight-core system, respectively. M. D. Gomony *et al*. [4] proposed a real-time multi-channel controller that could be feasibly applied in a high-definition video and graphics processing system.

Additionally, several hardware-based controllers have recently been presented. M. Vanegas *et al*. [5] described a multi-port memory controller (MPMC) with multiple abstract access ports to serve all transactions at the same time. A four-level controller hierarchy with time-division multiplex based arbiter for the H.264 1080p@30fps video decoder was proposed by Bonatto A. C. *et al*. [6]. T. Hussain *et al*. [7] designed a controller that accessed to memory by several defined patterns in order to reduce the access time. Two commercial MPMCs for high-bandwidth applications were also provided by Xilinx [8] and Altera [9]. A controller based on credit borrow and repay technique, which minimized the latency while preserving minimum bandwidth guarantees, was introduced by Zefu Dai *et al*. [10]. Our previous work [11, 12] focused on a parallel pipelining MPMC for multimedia applications, which achieved write and read bandwidth of 82 % and 87 %, respectively.

These mentioned works, however, still contain some disadvantages: (1) the hardware implementation of controllers is costly due to its complex architecture [2 - 4]; (2) the increase in number of access ports caused a negative effect on the total bandwidth utilization [5]; (3) the lack of support for general-purpose applications [6, 7]; (4) the reduction in latency is unconsidered [5 - 9], [11, 12]; (5) bandwidth efficiency seems insufficient for data-intensive applications [5 - 8, 10]. To address those problems, we propose an FPGA-based MPMC with advantages of flexibility, low latency, and high bandwidth, as summarized below.

*Flexibility*: depending on each specific application, the configuration parameters such as the number of granted ports, the burst count, and the access addresses can be configured in run-time. Moreover, any application system containing various operating clocks and data widths can easily connect to MPMC without adding extra interfaces.

*Latency*: dual-clock dual-port FIFOs (DCDWFFs) temporarily store the transfer data of application systems and allow users to put data in and get data out instantly if such data are available, thereby reducing the access latency. Moreover, the parallel and pipelining architecture are employed to minimize the latency at every processing stage of MPMC.

*Bandwidth*: the arbitration scheme, so-called window-based first-come-first-serve (WFCFS), is proposed to reduce the negative impact on the total bandwidth utilization and guarantee fair bandwidth distribution. Moreover, WFCFS architecture is fairly simple to implement in hardware.

The proposed MPMC is designed by Verilog HDL, simulated by Modelsim, and validated in a Terasic SoCKit development board [12], which contains an Altera Cyclone V FPGA and a 1-GB SDRAM DDR3. MPMC operates at 150 MHz and provides the theoretical bandwidth of 19.2 Gbps. It supports up to 32 parallel bidirectional ports that accept connections with different clocks and data widths. More significantly, the bandwidth utilization is approximately 93.2 % of the theoretical bandwidth, whereas the latency of each port is much smaller than that of other designs. The hardware resource at maximum settings only costs 4 % of lookup tables and 3 % of registers of a Cyclone V FPGA.

The remainder of this paper, then, is organized as follows. Section 2 describes in detail the hardware architecture of the proposed MPMC. Section 3 shows the experiment results validated in an FPGA under different settings. Section 4, finally, gives the conclusion and future works.

## 2. HARDWARE IMPLEMENTATION

### 2.1. Overview

The proposed MPMC is responsible for data transfer between an application system (APPSYS) and an external memory SDRAM, as depicted in Fig. 1. It consists of four main modules, namely INTERFACE, CONFIG, ARBITER, and PHY. The *N*-bidirectional-port INTERFACE keeps the temporary data to speed up the memory transactions. CONFIG stores all configuration parameters received from APPSYS in its internal registers. The key module ARBITER manages all data transactions based on the given parameters. The Altera PHY controls the physical layer of SDRAM interface, i.e. translates all requests from ARBITER into SDRAM commands and then transfers them reliably to SDRAM. The efficient architecture of the MPMC front-end, which includes INTERFACE, CONFIG, and ARBITER, is our primary focus in this paper.

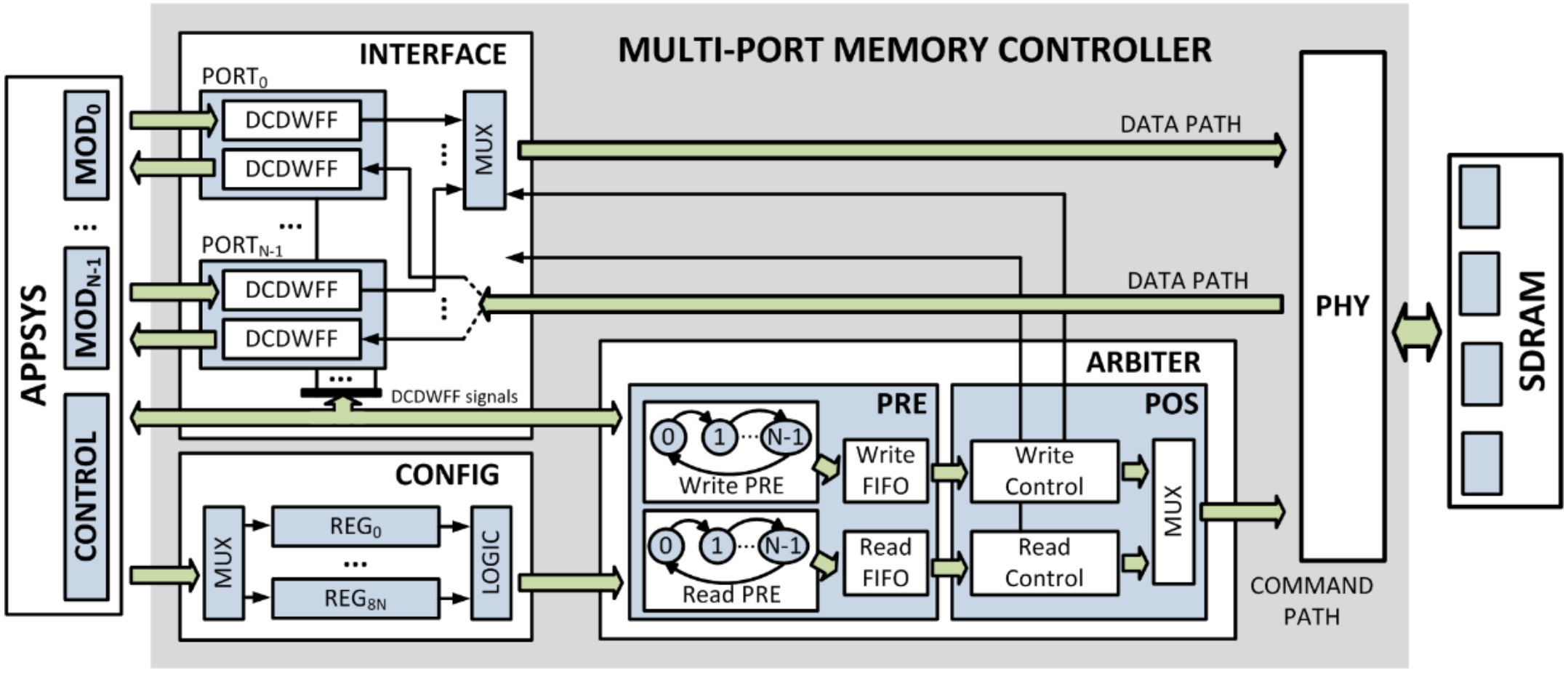

*Figure 1.* The general block diagram of the proposed MPMC.

#### *2.1.1. INTERFACE Module*

INTERFACE contains *N* PORTs and each one is composed of two DCDWFFs for read and write requests. Therefore, the access flexibility is improved since any multi-clock multi-port APPSYS can connect to MPMC. INTERFACE also guarantees robust transfers between MOD and PHY to minimize the problem of metastability, data loss, and data incoherency. Additionally, it minimizes the memory access latency, i.e., the time from a request being presented at MOD until it is processed completely, by using parallel DCDWFFs to separate the data path between MOD and PHY.

The architecture of a DCDWFF used in write requests is shown in Fig. 2. It includes two pairs of gray counters and shift registers, one dual-port memory, and one control unit. The write requests depend on two status signals, *full* and *almost_full*. In fact, if *full* is zero, write data *wr_data* are fed into DCDWFF together with the assertion of write enable *wr_en*. Otherwise, MOD waits until *full* turns into zero. As soon as DCDWFF keeps a certain amount of data, *almost_full* becomes one and then *rd_en* is asserted by ARBITER so that DCDWFF starts to transfer data *rd_q* to PHY.

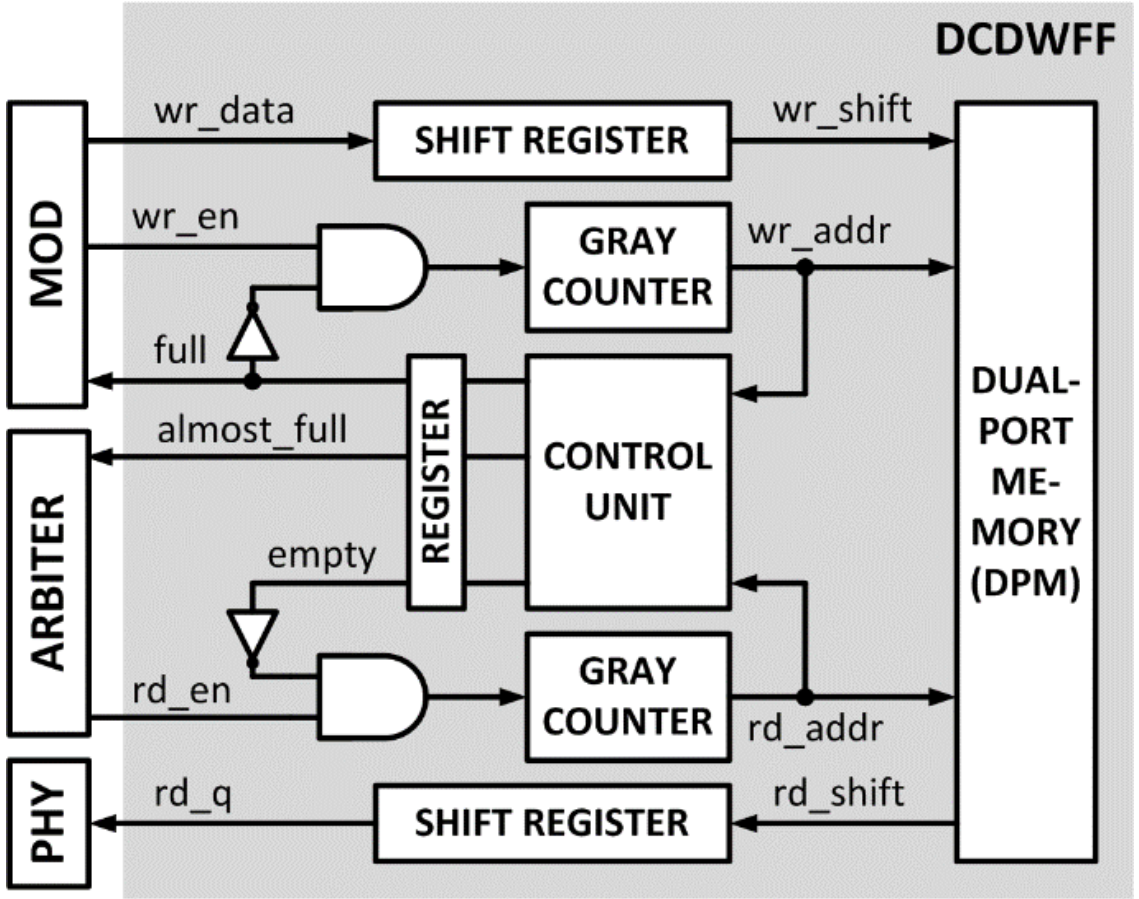

*Figure 2*. The hardware architecture of DCDWFF for write requests.

### *2.1.2. CONFIG Module*

CONFIG contains a set of registers to store the entire MPMC configuration as depicted in Fig. 3. Those registers include the number of used ports *N*, burst counts (BCs), and start/end/current addresses of transfers (SAs/EAs/CAs). The design supports *N* up to 32, BCs up to 64, and SAs/EAs/CAs up to four gigabytes. To improve the access flexibility, BCs, SAs, EAs, and CAs are separate for read and write requests. At the beginning of each transfer, APPSYS sequentially dispatches a set of configurations to the corresponding registers. During the operational process, ARBITER updates CAs by Eq. (1) and uses all given parameters to perform the scheduling.

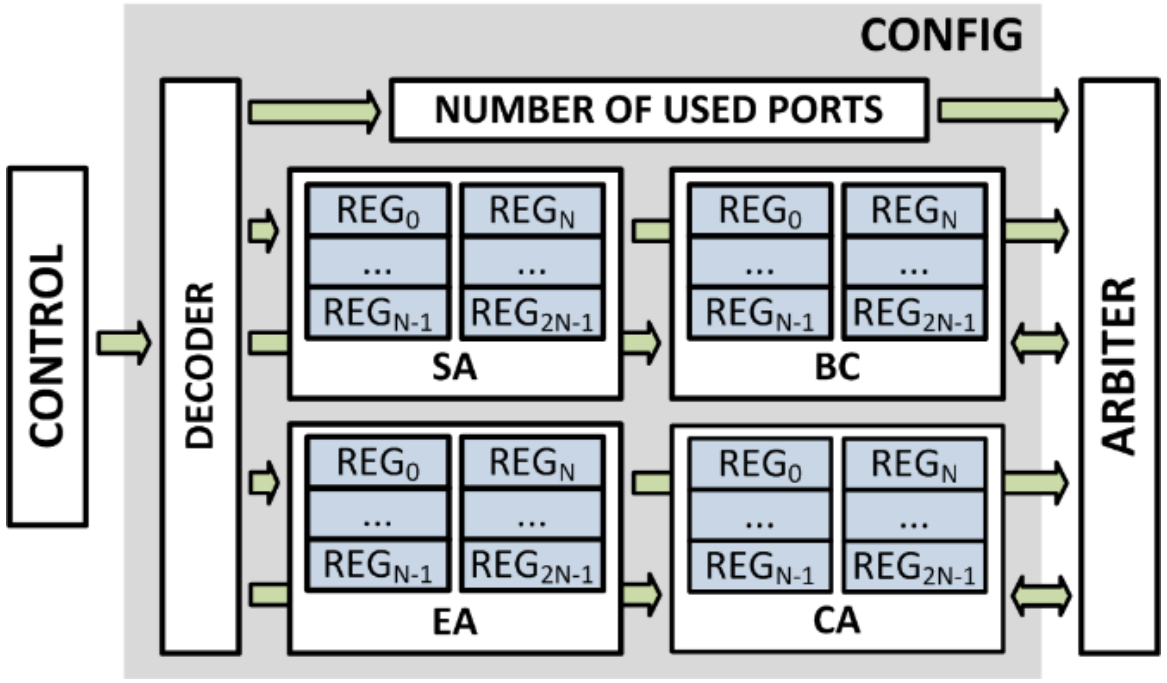

*Figure 3.* The general block diagram of CONFIG.

$$CA_i = \begin{cases} SA_i & \text{start of transfer} \\ CA_i + BC_i & \text{if} \quad \mathrm{CA}_i < \mathrm{EA}_i \end{cases} \quad (1)$$
$$\text{where } i \in \{0, N\text{-}1\}$$

In a multi-processor system where a memory region can be shared among several MODs, a bank conflict is likely to occur and cause a negative impact on bandwidth utilization. If there are two consecutive accesses to one bank, MPMC first sends the address to an SDRAM device, receives the data requested, and then waits for the SDRAM device to precharge and reactivate before initiating the next data transaction, thus wasted several clock cycles. To reduce the waiting clocks, bank assignment must be planned on in advance to exploit bank interleaving such as [13]. A basic example of the MOD-PORT-BANK assignment, in the case of $N$ = 4, is shown in Fig. 4. In Fig. 4(a), $PORT_0$ and $PORT_1$ access to $BANK_0$ consecutively, which obviously causes the bank conflict. However, in Fig. 4(b), the order of all accesses is $BANK_0$, $BANK_1$, $BANK_0$, and $BANK_2$, thereby eliminating the wasting clocks. These assignments are simply implemented by changing SA in CONFIG. The impact on bandwidth utilization of bank interleaving experiments is detailed in Section 3.

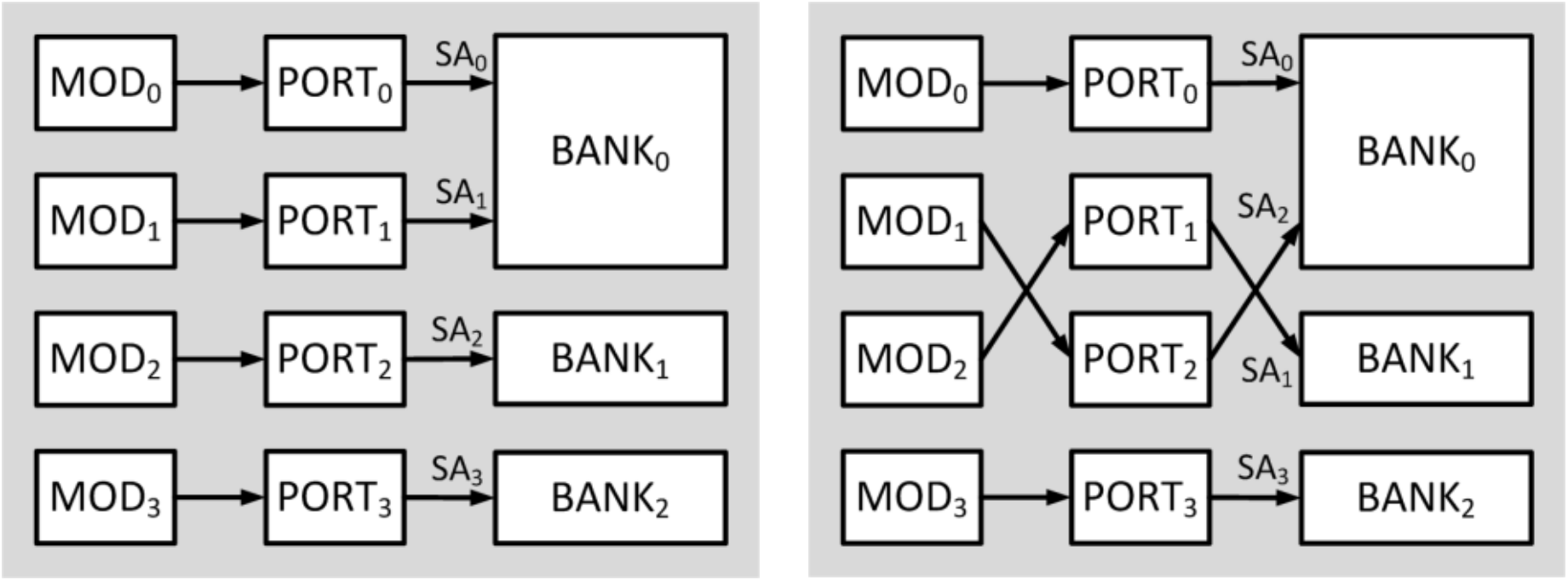

*Figure 4.* An example of accesses (a) without bank interleaving and (b) with bank interleaving.

### *2.1.3. ARBITER Module*

First-come-first-serve (FCFS) is an efficient scheduling process regarding performance and complexity [14]. However, in FCFS, a short-time request can get stuck behind long-time requests. In addition, FCFS may process a read/write request immediately after a write/read request, which causes several idle cycles on the SDRAM bus, so-called read/write turnaround. The first problem is solved by using DCDWFFs, i.e., data of incoming short-time requests are temporarily stored in DCDWFFs while current long-time requests are being processed. To overcome the second problem, we propose a window-based FCFS (WFCFS) arbitration scheme that effectively minimizes the number of read/write turnaround. Moreover, parallel and pipeline architecture are implemented to reduce processing time.

Figure 5 shows an example of WFCFS at $N$ = 4. Assume that read and write requests of $PORT_0$, $PORT_1$, $PORT_2$, and $PORT_3$ are labeled as $R_0$, $R_1$, $R_2$, and $R_3$ and $W_0$, $W_1$, $W_2$, and $W_3$, respectively. Furthermore, $BC_{W0}$, $BC_{W1}$, $BC_{W2}$, and $BC_{W3}$ are named as the BCs of correspondent write transactions. To begin with, ARBITER conducts a poll from $R_0$ to $W_3$. Because only $R_0$, $R_2$, and $R_3$ are ready at that moment, they are put into the read FIFO (RFF), and the window size becomes three, as shown in Fig. 5(a). Subsequently, the read control (RCTRL) sends all read requests with related parameters to PHY, as shown in Fig. 5(b). Simultaneously, since all requests $W_0$ to $W_3$ are ready, ARBITER puts all of them to write FIFO (WFF), and the window size becomes four. Afterwards, the write control (WCTRL) dispatches all write requests and data to PHY. Moreover, at the same time, the read data are returned to the corresponding ports since both RCTRL and WCTRL operate in parallel. The latency caused by read/write turnaround is reduced significantly due to the use of the windows. The impact on bandwidth utilization of WFCFS experiments is also in Section 3.

The hardware architecture of ARBITER is composed of two main modules, PRE and POS, as shown in Fig. 1. PRE checks whether a certain MOD requires the access to SDRAM and POS executes this request if the connecting PORT is available.

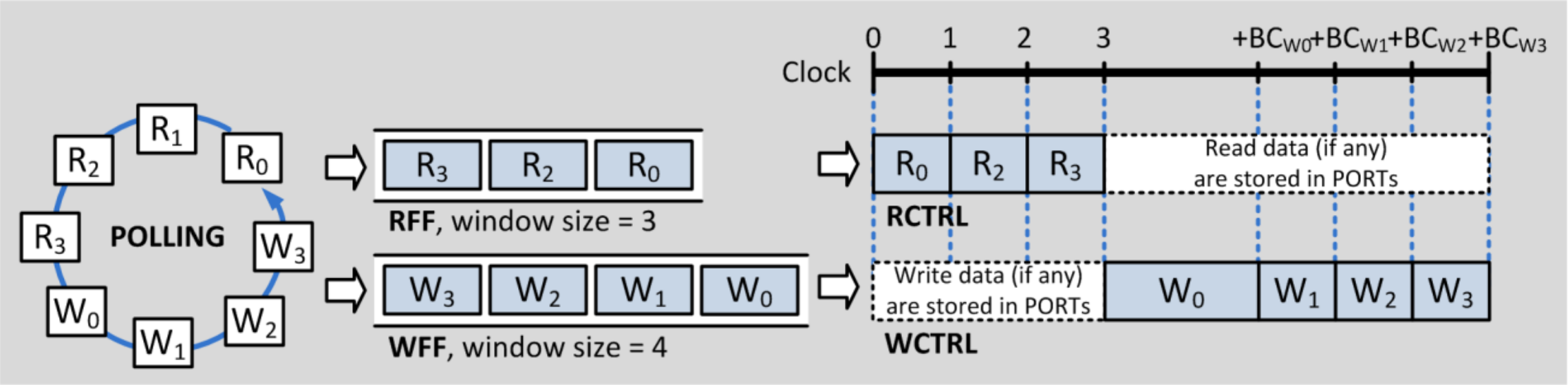

(a) (b)

*Figure 5.* An example of (a) PRE and (b) POS.

**PRE Module**: read and write requests independently by using a pair of sub-modules, so-called write PRE and read PRE, respectively. Each sub-module includes a POLLING circuit, a 32-bit FLAG register, and an RFF/WFF. The simplified architecture of write PRE is shown in Fig. 6. All components operate in pipelining to maximize the throughput.

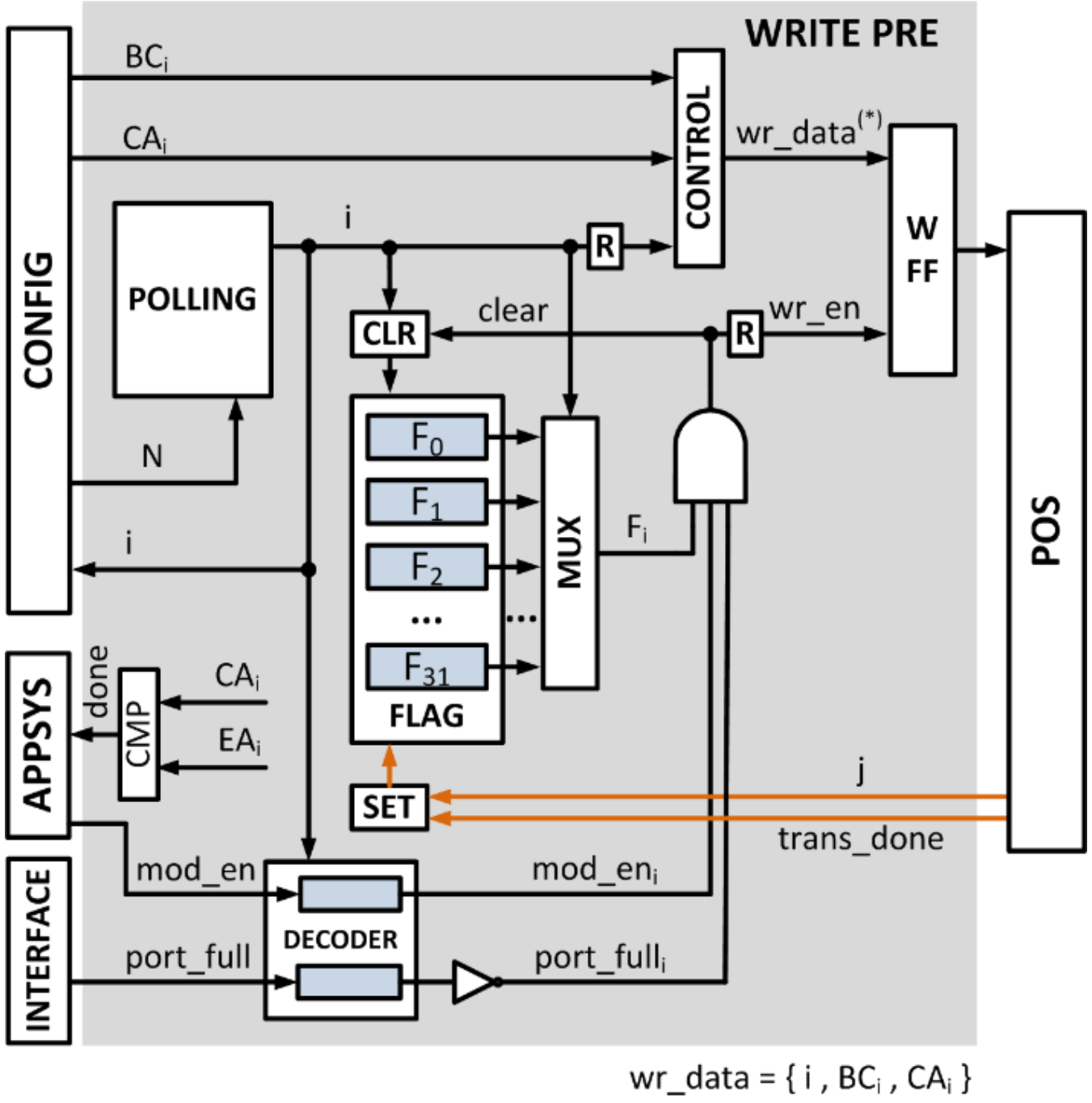

*Figure 6.* The hardware architecture of PRE for write requests.

Initially, the number of used ports $N$ is loaded to POLLING and all FLAG bits are set to high. The 32-bit *mod_en* shows the enabled MODs and the 32-bit *port_full* indicates the availability of ports. During the write transfers, POLLING scans each MOD for the request and then outputs the index of its connecting port $i$. The index is used to retrieve the port information stored in CONFIG, including CA, EA, and BC. Simultaneously, DECODER deploys $i$ to obtain $mod_en_i$ and $port_full_i$. If all three bits $F_i$, $mod_en_i$, and $port_full_i$ are one, $F_i$ is clear by CLR in the next clock. $F_i = 0$ indicates $PORT_i$ is in progress.

In the subsequent cycle, $EA_i$, $CA_i$, and $BC_i$ arrive to PRE. The combination of $i$, $BC_i$, and $CA_i$ are put into WFF with the assertion of write enable *wr_en*. Based on the received parameters, POS can write $BC_i$ data words to address $CA_i$ of SDRAM. Additionally, upon completing, the index $i$ is returned to PRE as $j$. SET uses $j$ and transaction done signal *trans_done* to turn $F_j$ into one, i.e., $PORT_j$ is ready for the next set of requests. Both set and clear process are performed simultaneously. If the transfer is completed, i.e., $CA_i \geq EA_i$, mod_eni turns into zero so that POLLING will not check $MOD_i$. Similarly, the read PRE shares the same architecture with the write PRE and RFF stores all parameters for the read process. Due to POLLING, bandwidth is distributed fairly among all requests.

**POS Module**: read and write requests independently by RCTRL and WCTRL. Each module includes three parallel and pipeline tasks so as to maximize throughput. Furthermore, each task is formed by several counters with logic circuits, instead of the finite state machines, to reduce hardware utilization.

The block diagram of WCTRL, as shown in Fig. 7(a), includes three tasks, namely $W_A$, $W_B$, and $W_C$. $W_A$ is responsible for retrieving the information of requests from PRE and returning the transaction done signal to PRE. Upon receiving those parameters, $W_B$ commands PORT to send data to PHY directly. $W_C$ monitors the indicators from PHY to end the transaction and signal to $W_A$. Similarly, RCTRL consists of three tasks, namely $R_A$, $R_B$, and $R_C$, as shown in Fig. 7(b). As soon as $R_A$ receives information, $R_B$ sends the commands to PHY. $R_C$ monitors the returned data and signal to $R_A$ if all data are buffered completely. It should be noted that read requests are considered as complete upon receipt of the first read data while the write requests are counted as complete if all write data are sent to PHY successfully.

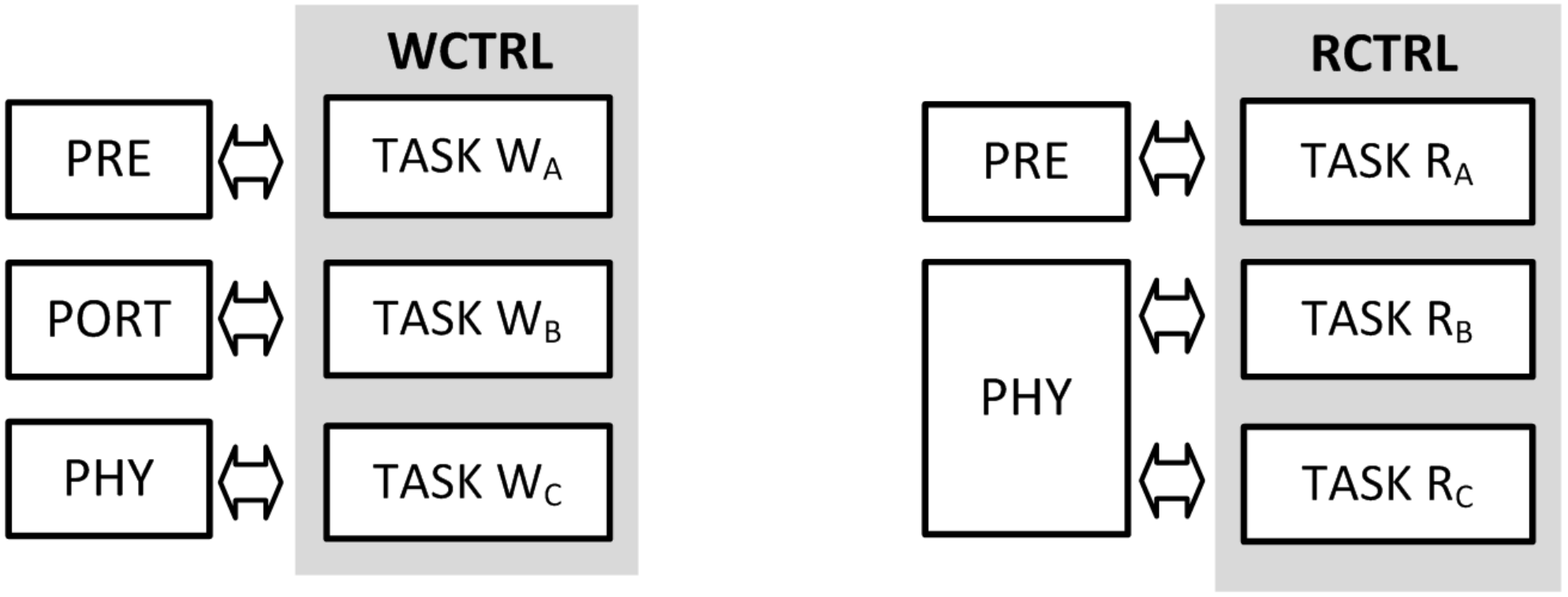

(a) (b)

*Figure 7*. The functionality of (a) WCTRL and (b) RCTRL.

## 3. PERFORMANCE ANALYSIS

In this section, we describe the experimental frameworks used to evaluate the performance of an mpmc concerning bandwidth utilization, access latency, and resource consumption, as compared to other designs.

Bank interleaving (BKIG) can improve BW efficiency by mapping each port to the memory bank appropriately. To evaluate such improvements, we conducted three experiments namely EXPA, EXPB, and EXPC with the bank assignments shown in Table 1. In EXPA, all MODs only access $BANK_0$. In EXPB, $MOD_0$ and $MOD_2$ access $BANK_0$ while the rest accesses $BANK_1$. In EXPC, every MOD is assigned to a different bank. Fig. 8 makes the comparison of EFF among three experiments at BC = {4, 8, 16, 32, 64}. EXPC always provides the highest EFF because the bus turnaround time is long enough for MPMC to ideally send one data request to each of the banks in consecutive clock cycles. In fact, one bank undergoes its precharge or activate cycle while another is being accessed. EXPB attains nearly the same BW as EXPA at BC = {32, 64}. However, its BW is significantly reduced at lower BCs because of the insufficient bus turnaround time. EXPA shows the worst BW as a result of bank conflict. Therefore, depending on the particular applications, the MOD-PORT-BANK assignment must be planned on in advance.

*Table 1.* The bank assignment in three experiments.

| | $PORT_0$ | $PORT_1$ | $PORT_2$ | $PORT_3$ |
|---|---|---|---|---|
| EXPA | $BANK_0$ | $BANK_0$ | $BANK_0$ | $BANK_0$ |
| EXPB | $BANK_0$ | $BANK_1$ | $BANK_0$ | $BANK_1$ |
| EXPC | $BANK_0$ | $BANK_1$ | $BANK_2$ | $BANK_3$ |

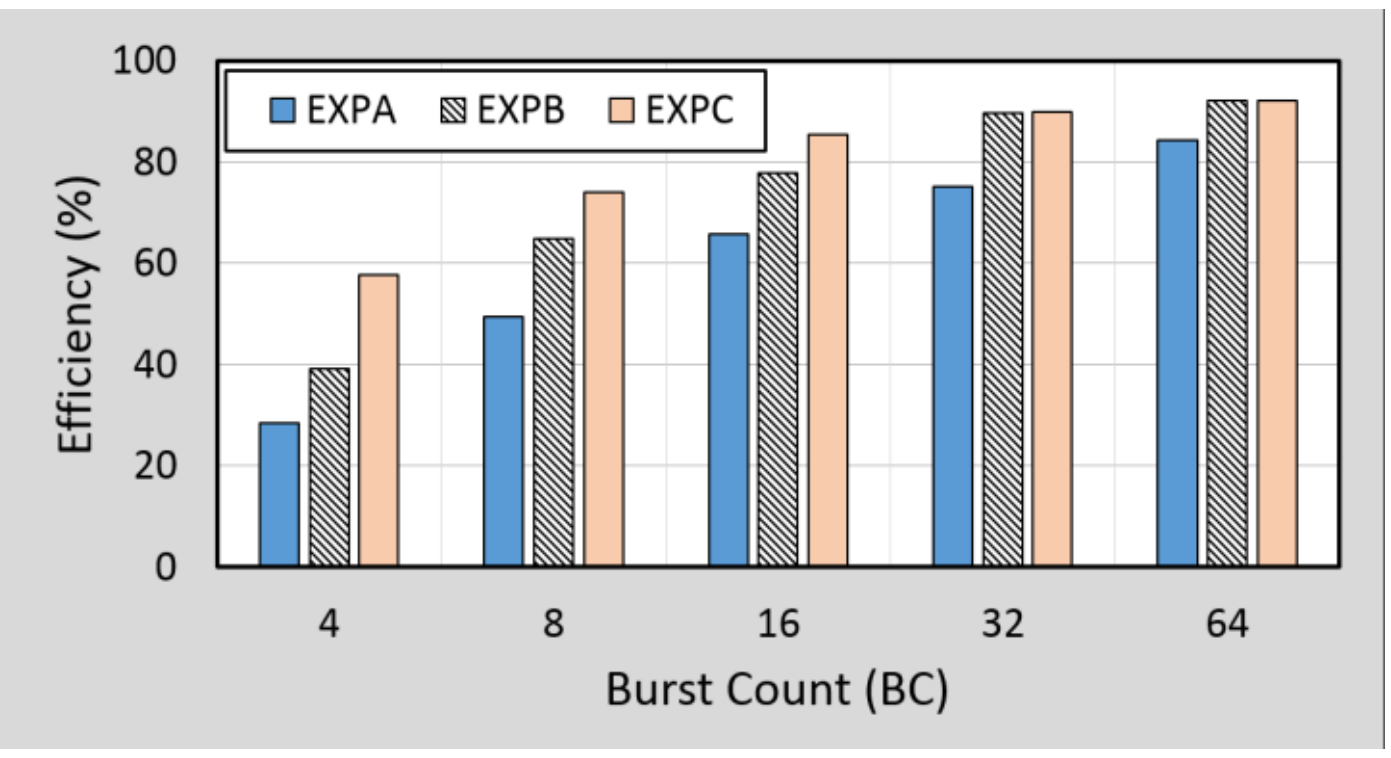

*Figure 8.* The comparison of bandwidth utilization among three experiments.

The proposed arbitration scheme WFCFS allows ARBITER to keep the requests in WFF and RFF and then process several of them each time, whereas FCFS executes each request immediately upon receiving it. To assess the performance of WFCFS, we conducted an experiment EXPD that only deploys FCFS and compared its BW with that from EXPC above. The window size varies up to four due to the number of used ports, $N$ = 4. According to Fig. 9, EFF of EXPC is always higher than that of EXPD since WFCFS minimizes the read/write turnaround effectively. Moreover, the higher BC can ease the loss of BW, i.e. EFF of EXPD reduces around 5 % at BC = 64 to 17 % at BC = 4 as compared to EFF of MPMC. In short, exploiting both BKIG and WFRFS could improve EFF, at least, 12.9 % in case of $N = 4$.

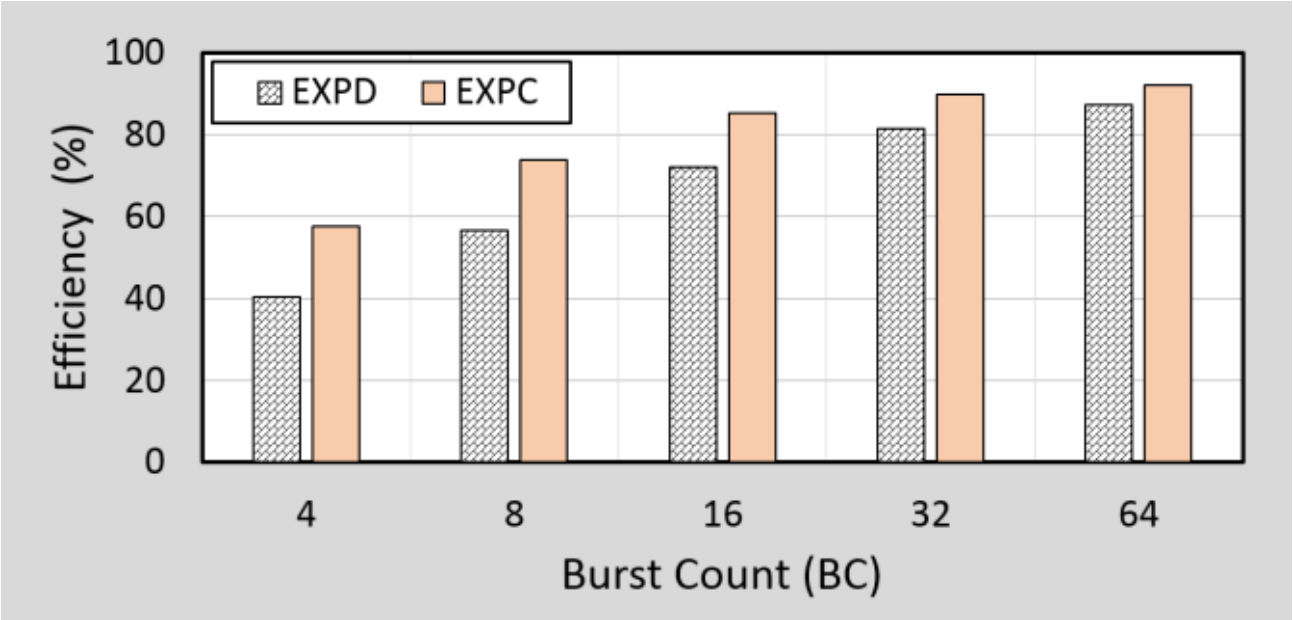

*Figure 9.* The comparison of bandwidth utilization between EXPC using WFCFS and EXPD using FCFS.

The peak BW is measured by performing continuous requests from all MODs to MPMC. In addition, two consecutive PORTs access to two different banks to exploit BKIG. Fig. 10 illustrates the achieved BW at BC = {4, 8, 16, 32, 64} and the number of used ports $N = \{2, 4, 8, 16, 32\}$. The horizontal axis represents BC while the vertical axis represents BW. It appears the total BW counts on both $N$ and BC. Actually, BW increases with $N$ because at larger $N$, POS can process more commands stored in RFF and WFF at each time. Furthermore, BW increases with BC because each column-access command at larger BC can transfer more data per each transaction. BW reaches the maximum value of 17.9 Gbps, or EFF of 93.2 %, at $N = 32$ and BC = 64. Furthermore, BW is distributed equally among ports since each of them uses the same configuration.

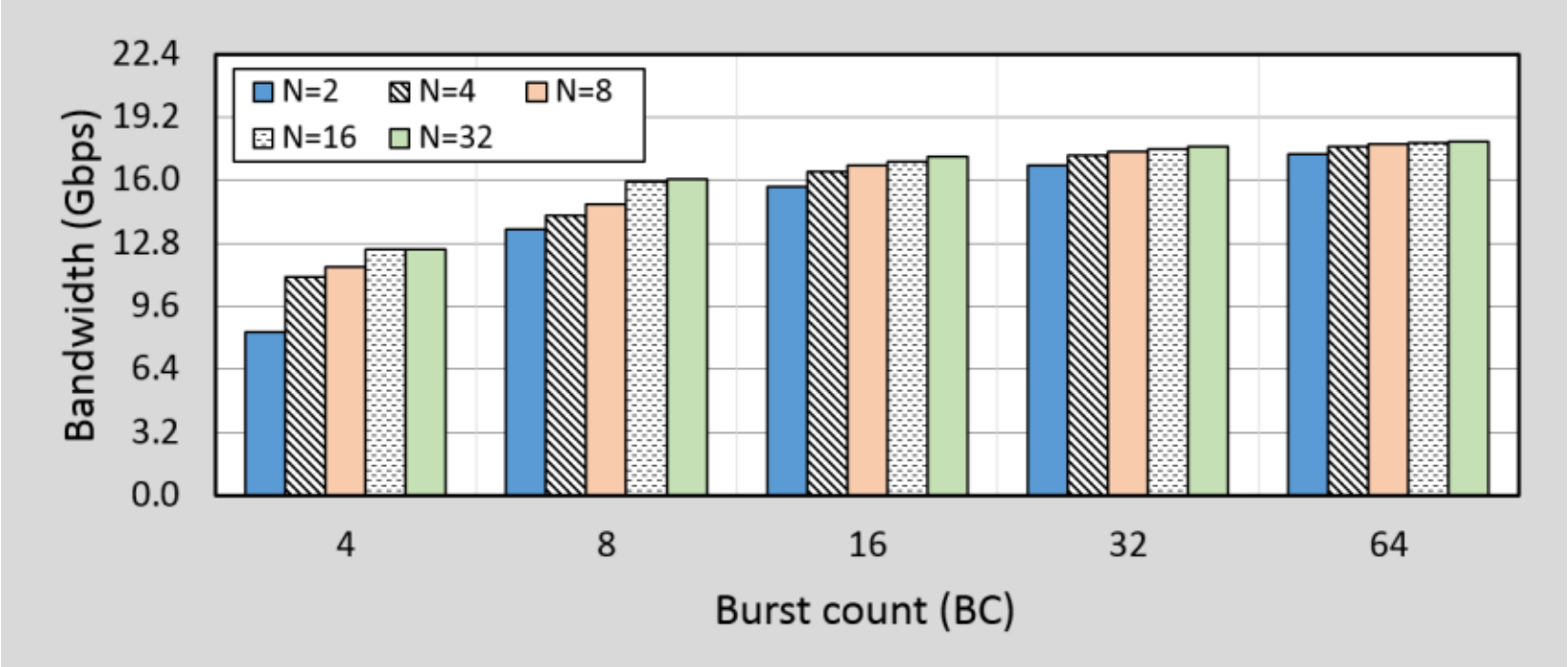

*Figure 10.* The comparison of bandwidth utilization of MPMC at different N and BC.

The effect of $N$ on BW between our MPMC with a design DESA [5] is shown in Fig. 11. In DESA, as $N$ increases, BW on each port reduces significantly, which leads to the reduction of total BW. Assuming an EFF of 100 % occurs at the highest BW, DESA achieved such BW at $N = 2$ and BW drastically declined nearly 60 % until $N = 10$. On the contrary, in our design, the total BW reaches the maximum at higher $N = 10$ and slightly reduces by around 2 % at $N = 2$. Although DESA supports many kinds of memory chips such as DDR, DDR2, and SSRAM, its BW reduction is quite difficult for data-intensive applications.

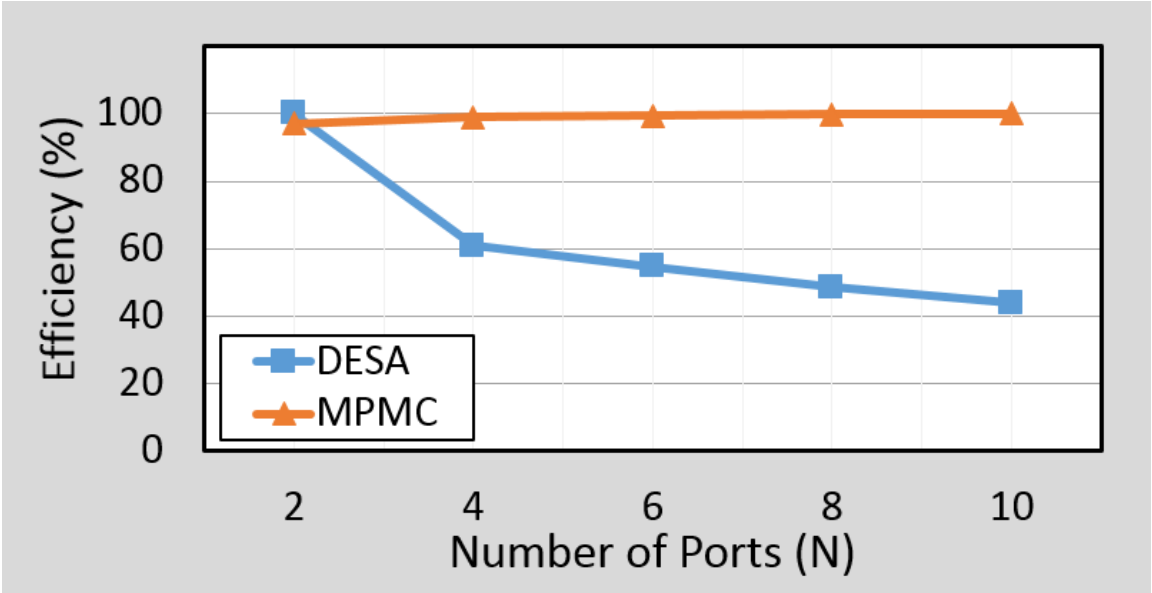

*Figure 11.* The comparison of bandwidth loss between two designs as N increases.

The comparison of EFF between the proposed MPMC with three other works DESB, DESC, and DESD are illustrated in Fig. 12. The information from FPGA and SDRAM of each design is summarized in Table 2. The EFF of write requests and read requests are analyzed independently. Figure 12 illustrates EFF at $N = \{2, 4, 8\}$ and BC = {16, 32, 64}. Generally, EFF of write requests is lower than EFF of read requests since in writing, MPMC must read the entire row, and then write back the old data along with the data required to write. Thus, the procedure for writing data to the array includes both read and write processes. In our MPMC, write request and read request achieve EFF of 92.2 % and 94.8 %, respectively.

*Table 2.* The parameters of SDRAM in compared designs.

| | SRAM DDR3 | Theoretical bandwidth utilization (Gbps) |
|---|---|---|
| DESB [8] | 400 MHz, 32 bits | 25.6 |
| DESC [8] | 400 MHz, 16 bits | 12.8 |
| DESD [9] | 300 MHz, 32 bits | 19.2 |
| MPMC | 300 MHz, 32 bits | 19.2 |

*Table 3.* A comparison of resource utilization among several designs.

| | Device | N | LUTs | REGs |
|---|---|---|---|---|
| DESE **[6]** | Xilinx Virtex-5 | 3 | 2,739 | 2,714 |
| DESF **[7]** | Xilinx Virtex-5 | 4 | 3,971 | 2,883 |
| DESB **[8]** | Xilinx Virtex-6 | 8 | 3,600 | 5,860 |
| DESA **[5]** | Xilinx Virtex-4 | 10 | 1,733 | - |
| DESD **[9]** | Altera Stratix IV | 16 | 4,221 | 2,424 |
| MPMC_2 | Altera Cyclone V | 2 | 1,251 (1 %) | 1,804 (1 %) |
| MPMC_4 | Altera Cyclone V | 4 | 1,322 (1 %) | 2,053 (1 %) |
| MPMC_8 | Altera Cyclone V | 8 | 1,768 (1 %) | 2,504 (1 %) |
| MPMC_16 | Altera Cyclone V | 16 | 2,634 (2 %) | 3,679 (2 %) |
| MPMC_32 | Altera Cyclone V | 32 | 4,453 (4 %) | 6,046 (3%) |

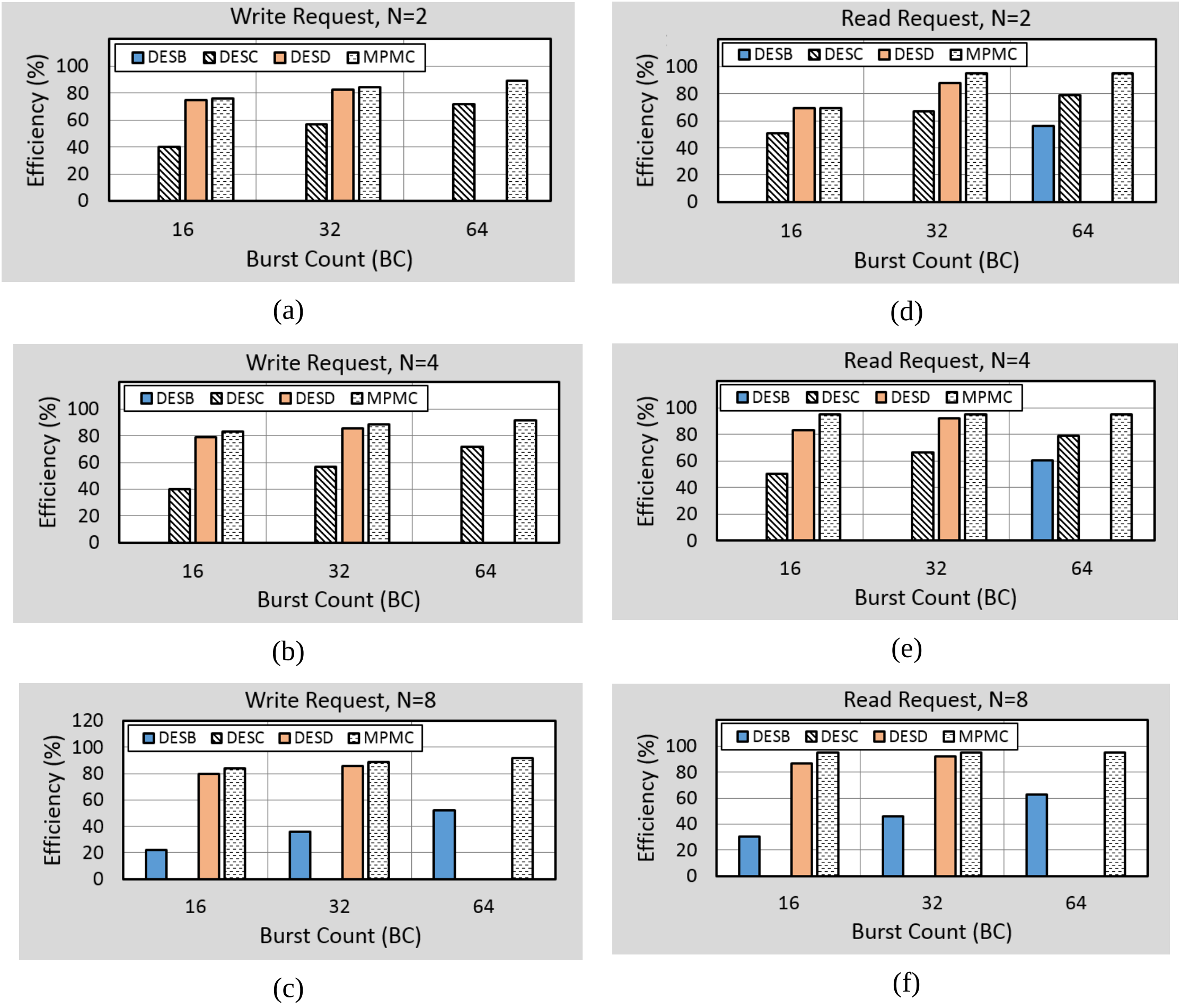

*Figure 12.* The comparison of bandwidth utilization among four designs at write process (a), (b), (c) and read process (d), (e), (f).

Table 3 draws the comparison of utilized lookup tables (LUTs) and registers (REGs) between our MPMC with the others at different *N*. The memory bits are not mentioned in this comparison because they depend on the application requirements. Suppose that MPMC_N indicates the resource of MPMC with *N* used port. In our design, both ARBITER and CONFIG cost around 700 of LUTs and 1,400 of REGs, and are independent of *N*. Furthermore, if more ports are utilized, LUTs and REGs increase correspondingly. It should be noted the design DESE, DESF, and DESA utilize unidirectional ports, i.e. read or write port only, while the others support bidirectional ports. In comparison with DESB and DESD, MPMC_8 and MPMC_16 cost more REGs to store the configuration parameters. At maximum settings of MPMC_32, we cost approximately 4 % of LUTs and 3 % of REGs.

## 4. CONCLUSIONS

In this study, we presented a configurable MPMC for high-bandwidth and low-latency applications. A pair of DCDWFFs is deployed in every port to minimize the access latency and allow the connection from any multi-clock multi-data-width APPSYS without adding extra interface resources. A WFCFS arbitration scheme with parallel pipelining architecture was proposed to improve the bandwidth utilization. The experimental results in an Altera Cyclone V FPGA prove that MPMC is fully operational with 32 concurrent connections at various clocks and data widths. The bandwidth efficiency at maximum settings is approximately 93.2 % and the access latency is significantly reduced as compared to other designs. Finally, the proposed MPMC has proven its performance in a data analytics system [17 - 19], where flexible access patterns and high-bandwidth utilization play a major role.